\def\fr#1#2{\hbox{${#1\over #2}$}}

\def\ni{\noindent}

\def\+{{(+)}}  \def\-{ {(-)} }   \def\0{ {(0)} }
\def\1{ {(1)} }  \def\2{ {(2)} }
\def\pd{\partial}

\def\con{\omega}

\def\sq{Q\kern-6pt/}
\def\sQ{Q\kern-12pt\nearrow}
\documentclass[11pt,eqs]{article}
\usepackage{latexsym,graphicx}

\def\be{\begin{equation}}             \def\ee{\end{equation}}
\def\ba{\begin{array}{rcl}}           \def\ea{\end{array}}
\def\beqa{\begin{eqnarray} }          \def\eeqa{\end{eqnarray} }
\def\beqalign{\begin{eqalign}}        \def\eeqalign{\end{eqalign}}
\def\leq#1{\label{eq:#1}}             \def\eq#1{(\ref{eq:#1})}
\def\bsubeq{\begin{subequations}}     \def\esubeq{\end{subequations}}
\def\bitem{\begin{itemize}}           \def\eitem{\end{itemize}}

\def\DJ{\leavevmode\setbox0=\hbox{D}\kern0pt
 \rlap{\kern.04em\raise.188\ht0\hbox{-}}D}
\def\dj{\leavevmode\setbox0=\hbox{d}\kern0pt
 \rlap{\kern.215em\raise.46\ht0\hbox{-}}d}

\newcommand{\bd}{\begin{displaymath}}
\newcommand{\ed}{\end{displaymath}}

\begin{document}
\title{CT-duality as a local property of the world-sheet
\thanks{Work supported in part by the Serbian Ministry of Science and
Environmental Protection, under contract No. 1486.}}
\author{ B. Sazdovi\'c\thanks{e-mail address:sazdovic@phy.bg.ac.yu}\\
{\it Institute of Physics, 11001 Belgrade, P.O.Box 57, Yugoslavia} }
\date{}
\maketitle

\begin{abstract}
In the present article, we study the local features of the world-sheet in the
case when probe bosonic string moves in antisymmetric background field. We
generalize the geometry of surfaces embedded in space-time to the case when
the torsion is present. We define the mean extrinsic curvature for spaces with
Minkowski signature and introduce the concept of mean torsion. Its orthogonal
projection defines the dual mean extrinsic curvature. In this language, the
field equation is just the equality of mean extrinsic curvature and extrinsic
mean torsion, which we call CT-duality. To the world-sheet described by this
relation we will refer as CT-dual surface.
\end{abstract}

\ni {\it Keywords \/}: world-sheet, torsion, CT-duality \par

\ni {\it PACS number(s)\/}: 02.40.-k,  11.25.-w,  04.90.+e \par

\section{Introduction}
\setcounter{equation}{0}

In general relativity, the action for the point particle is proportional to
the length between initial and final points. For the geodesic line, which is
solution of the equation of motion, the action is minimal. On the other hand,
there exists a definition of the geodesic in term of local properties of the
trajectory in the neighborhood of every given point. This definition is
equivalent to the previous one, because it produces the same equation of
motion. According to local features, geodesic is self-parallel line, which
means that tangent vector after parallel transport along this line remains
tangent.

In the string theory the action is area of the world-sheet between given
initial and final positions of the string. The solution of the equations of
motion, by definition, is minimal surface. What is definition of the  minimal
surface in term of local properties of the surface in the neighborhood of
every given point? The direct generalization of "self-parallel" surface is
impossible, because even initial and final positions of the string are not
necessary self-parallel. The local features of minimal surface is known in the
literature and this is the condition that all mean extrinsic curvatures vanish
\cite{BN,Sor}.

We are interested in  more general case, when the string
propagates in nontrivial massless background, which beside metric
includes also antisymmetric tensor. The corresponding Lagrangian
of the theory has been used in the literature
\cite{BS3}-\cite{BNS}, in order to derive the classical space-time
equations of motion from the world-sheet quantum conformal
invariance. The goal of the present paper is to describe the
world-sheet equations of motion in terms of local features, in the
presents of antisymmetric field.

In Sec. 2, we formulate the bosonic string theory which we are going to
consider, and shortly repeat some results of ref. \cite{BS1}, which we will
need later.

In Sec. 3, we consider general theory of surfaces embedded into Riemann-Cartan
space-time. We define the induced world-sheet variables: metric tensor and
connection, and extrinsic one: second fundamental form (SFF). The world-sheet
tangent vector, after parallel transport along world-sheet line with
space-time connection, is not necessarily a tangent vector. Its world-sheet
projection defines the induced connection and its normal projection defines
the SFF.

In Secs. 4 and 5. we explain the local properties of the world-sheet in the
presence of background fields. First we consider the case in the absence of
antisymmetric field. We precisely define mean extrinsic curvature (MEC) in
Minkowski space-time, and support the result of refs.\cite{BN,Sor}.

We generalized the concept of the SFF and of the MEC to the case where the
space-time has nontrivial torsion.  We define the mean torsion, ${}^\circ
T^\mu$, and show that its orthogonal projection, ${}^\circ T_i$, is dual mean
extrinsic curvature (DMEC), ${}^\circ T_i={}^\ast H_i$. This enables us to
introduce the self-dual (self-antidual) condition, ${}^\circ H_i = \pm {}^\ast
H_i$, to which we will refer as CT-duality. Finally, we prove the main result,
that the field equations in the presence of antisymmetric field have the form
of CT-duality and that according to the local properties the world-sheet is
self-dual surface.

Appendix A is devoted to the world-sheet geometry.

\section{Formulation of the theory}
\setcounter{equation}{0}

Let us consider the action
\cite{GSW}-\cite{BNS}
\be
S= \kappa  \int_\Sigma d^2 \xi \sqrt{-g} \left[ {1 \over
2}g^{\alpha\beta}G_{\mu\nu}(x) +{\varepsilon^{\alpha\beta} \over \sqrt{-g}}
B_{\mu\nu}(x)\right] \partial_\alpha x^\mu \partial_\beta x^\nu
 \, ,   \leq{ac}
\ee
which describes bosonic string propagation in $x^\mu$-dependent background
fields: metric $G_{\mu \nu}$ and antisymmetric tensor field $B_{\mu\nu}=-
B_{\nu\mu}$. Let $x^\mu(\xi) \,  (\mu =0,1,...,D-1)$ be the coordinates of the
$D$ dimensional space-time $M_D$ and $\xi^\alpha \, (\xi^0 =\tau ,
\xi^1=\sigma)$ the coordinates of two dimensional world-sheet $\Sigma$,
spanned by the string. The corresponding derivatives we will denote as
$\partial_\mu \equiv {\partial \over \partial x^\mu}$ and $\partial_\alpha
\equiv {\partial \over
\partial \xi^\alpha}$ and the intrinsic world-sheet metric by  $g_{\alpha
\beta}$.

We will briefly review some results of the ref. \cite{BS1},
adapted for the present case without dilaton field. It
is useful to define the currents
\be
j_{\pm \mu} =\pi_\mu +2\kappa \Pi_{\pm \mu \nu} {x^\nu}' \,  ,
\qquad  \Pi_{\pm \mu \nu} \equiv  B_{\mu \nu} \pm {1 \over 2}
G_{\mu \nu} \,  , \leq{jmi}
\ee
where $\pi_\mu$ is canonical momentum of the coordinate $x^\mu$.

The $\tau$ and $\sigma$ derivatives of the coordinate  $x^\mu$ can be
expressed in terms of the corresponding currents
\be
{\dot x}^\mu =  {G^{\mu \nu} \over 2 \kappa} (h^- j_{- \nu} - h^+ j_{+ \nu}) \,  ,
\qquad
x^{\mu \prime}= {G^{\mu \nu} \over 2 \kappa} (j_{+ \nu} - j_{- \nu})  \,  , \leq{xt}
\ee
where the components of the intrinsic metric tensor, $h^\pm$, are define in
the App. A.

The canonical Hamiltonian density, ${\cal H}_c= h^- T_- + h^+ T_+
$, and the energy momentum tensor components
\be
T_\pm =\mp {1 \over 4\kappa} G^{\mu\nu} j_{\pm \mu} j_{\pm \nu}  \,   ,  \leq{emtpm}
\ee
have the standard forms.
The equations of motion, for the action \eq{ac} is
\be
[x^\mu]  \equiv  \nabla_- \partial_+  x^\mu + \Gamma_{- \rho \sigma}^\mu
\partial_+ x^\rho \partial_- x^\sigma =0  \,  ,    \leq{lJ}
\ee
\be
[h^\pm]  \equiv  G_{\mu \nu}  \partial_\pm  x^\mu \partial_\pm x^\nu = 0   \,  ,  \leq{lh}
\ee
where the world-sheet covariant derivatives, $\nabla_\pm$, are defined in \eq{wscd}.
The expression in the $[x^\mu]$ equation is of the form
\be
\Gamma^\rho_{\pm \nu \mu}= \Gamma^\rho_{\nu \mu} \pm B^\rho_{\nu \mu}   \,  ,  \leq{cdc}
\ee
where
\be
B_{\mu \nu \rho}= \partial_\mu B_{\nu \rho} + \partial_\nu B_{\rho \mu} +
\partial_\rho B_{\mu \nu}= D_\mu B_{\nu \rho} + D_\nu B_{\rho \mu} + D_\rho
B_{\mu \nu}   \,  ,       \leq{fsB}
\ee
is the field strength of the antisymmetric tensor. It is a generalized
connection, which full geometrical interpretation has been investigated
in \cite{BS3}. Under space-time general coordinate transformations
the expression $\Gamma^\rho_{\pm \nu \mu}$ transforms as a connection.

As a consequence of the symmetry relations $\Gamma_{\mp \rho \sigma}^\mu =
\Gamma_{\pm \sigma \rho}^\mu$, we can rewrite eq. \eq{lJ} in the form
$[x^\mu]  \equiv  \nabla_+ \partial_-  x^\mu + \Gamma_{+ \rho \sigma}^\mu
\partial_- x^\rho \partial_+ x^\sigma =0$. So, all considerations we can also
apply to $\Gamma_{+ \rho \sigma}^\mu$.

\section{Geometry of surfaces embedded in Riemann-Cartan space-time}
\setcounter{equation}{0}

The geometry of surfaces, when the world-sheet is embedded in curved
space-time, has been investigated in the literature, see \cite{BN,Sor}. In
this section we will generalize these results for the space-times with
nontrivial torsion.

\subsection{Riemann-Cartan geometry}
\setcounter{equation}{0}

Let us first fix notation and shortly repeat some definitions of
refs. \cite{BS3}, \cite{MB} and \cite{HMMN}.

The affine linear connection, ${}^\circ \Gamma_{\rho \sigma}^\mu$, defines the
rule for parallel transport of the vector $V^\mu (x)$, from the point $x$ to
the point $x+dx$, as $V^\mu(x) \to   {}^\circ V^\mu_\parallel = V^\mu +
{}^\circ \delta V^\mu$, where
\be
{}^\circ \delta V^\mu = - {}^\circ \Gamma_{\rho \sigma}^\mu V^\rho   d
x^\sigma  \,    .   \leq{ptcn}
\ee
The covariant derivative is defined as
\be
{}^\circ D V^\mu = V^\mu (x+dx) - {}^\circ V^\mu_\parallel
\equiv  {}^\circ D_\nu V^\mu d x^\nu     \,   ,
\ee
where ${}^\circ D_\nu V^\mu = \partial_\nu V^\mu + {}^\circ \Gamma_{\rho
\nu}^\mu V^\rho$.

The connection is not necessary symmetric in the lower indices, and its
antisymmetric part is the torsion
\be
{}^\circ T^\rho_{\mu \nu} ={}^\circ \Gamma^\rho_{\mu \nu} - {}^\circ \Gamma^\rho_{\nu \mu} \,  .
\ee
It has a simple geometrical interpretation which we will need later. Let us
perform parallel transport of the unit tangent vectors $ t^\mu_1 \, (
t^\mu_2)$ along geodesics $\ell_2 \,  (\ell_1)$ at the distances $ d \ell_2 \,
(d \ell_1)$, respectively. The final vectors we denote by $\tilde{t}^\mu_1 \,
(\tilde{t}^\mu_2)$. They define directions of the geodesics $\tilde{\ell}_1 \,
(\tilde{\ell}_2)$, which ends at the points $D_2 \,  (D_1)$, at the
distances $d \ell_1 \,  (d \ell_2)$,  (Fig. 1). The
difference of the coordinates at the points $D_2$ and $D_1$ is proportional to
the torsion

\begin{figure}
\begin{center}
\includegraphics[height=6cm]{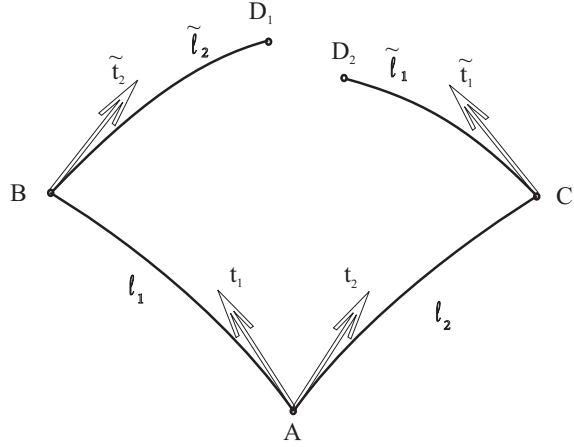}
\end{center}
\vspace*{-.5cm} \caption{{\it Geometrical meaning of the torsion.} }%
\end{figure}

\be
x^\mu(D_2) - x^\mu(D_1)  = {}^\circ T^\mu{}_{\rho \sigma}\, t_1^\rho \, t_2^\sigma \, d \ell_1 \, d \ell_2    \,   .       \leq{cod}
\ee
In fact {\it the torsion measures the non-closure of the "rectangle" $ABCD$}.

In the present case, the decomposition of the connection has a form
\be
{}^\circ \Gamma_{\mu ,\rho \sigma} = \Gamma_{\mu ,\rho \sigma} + {}^\circ K_{\mu \rho \sigma}
  \,   .     \leq{cde}
\ee
The first term is the Christoffel connection, $\Gamma_{\mu , \rho \sigma} = \frac{1}{2}
\partial_{ \{ \mu } G_{ \rho \sigma \} }$, and the second one is the contortion,
${}^\circ K_{\mu \rho \sigma} = \frac{1}{2} {}^\circ T_{ \{ \sigma \mu \rho \} }$,
where we introduce the Schouten braces according to
the relation $\{ \mu \rho \sigma \} = \sigma \mu \rho + \rho \sigma \mu - \mu
\rho \sigma$.

The first term is symmetric in $\rho, \sigma$ indices. In the second term we
can separate symmetric and antisymmetric parts, ${}^\circ K_{\mu \rho \sigma}=
{}^\circ K_{\mu (\rho \sigma)}+ \frac{1}{2} {}^\circ T_{\mu \rho \sigma}$,
where he symmetric part of the arbitrary tensor $X_{\mu \nu}$ we denote as
$X_{(\mu \nu)} \equiv \fr{1}{2}(X_{\mu \nu} + X_{\nu \mu})$. Consequently, we
have
\be
{}^\circ \Gamma^\mu_{\rho \sigma} = {}^\circ \Gamma^\mu_{(\rho
\sigma)} + \frac{1}{2} {}^\circ T^\mu_{\rho \sigma} \,  . \leq{c2}
\ee
The antisymmetric part of the connection \eq{cdc} is the Rieman-Cartan
torsion, seen by the string
\be
T_\pm{}^\rho_{\mu \nu} = \Gamma_\pm{}^\rho_{\mu \nu} -  \Gamma_\pm{}^\rho_{\nu \mu} =
\pm 2   B^\rho_{\mu \nu}      \,  .   \leq{T}
\ee
It is proportional to the field strength of the antisymmetric tensor field
$B_{\mu \nu}$. In this case the contortion is proportional to the torsion
$K_{\pm \mu \nu \rho} = \fr{1}{2} T_{\pm \mu \nu \rho} = \pm B_{\mu \nu \rho}
$. So, the expressions  \eq{cde} and \eq{c2} turn to \eq{cdc}.

\subsection{Induced and extrinsic geometry}

We will use the local space-time basis, relating with the coordinate one by
the vielbein $E^\mu_A = \{\partial_\alpha x^\mu , n_i^\mu \}$. Here,
$\partial_\alpha x^\mu = \{ {\dot x^\mu} , x'^\mu \}$ is local world-sheet
basis and $n^\mu_i \, (i=2,3,...,D-1)$ are local unit vectors, normal to the
world-sheet.

The world-sheet {\bf induced metric tensor}
\be
G_{\alpha \beta} = G_{\mu \nu} \partial_\alpha x^\mu
\partial_\beta x^\nu \,  ,  \leq{imt}
\ee
is defined by the requirement, that any world-sheet interval has the same
length, measured by the target space metric or by the induced one.

Similarly, the induced metric of a $D-2$ dimensional space, normal to the
world-sheet, is $G_{i j} = G_{\mu \nu}  n_i^\mu n_j^\nu$. The mixed induced
metric tensor $G_{\alpha i} = G_{\mu \nu} \partial_\alpha x^\mu n_i^\nu$,
vanishes by definition.

The world-sheet projection and the orthogonal projection of the arbitrary
space-time covector $V_\mu$, we will denote as
\be
v_\alpha = \partial_\alpha x^\mu V_\mu \,  , \qquad
v_i = n_i^\mu V_\mu     \,   .      \leq{pr}
\ee
In the space-time basis, tangent and normal vectors to the world-sheet
$\Sigma$ can be expressed respectively as
\be
V^\mu_\Sigma =  \partial_\alpha x^\mu v^\alpha  \,   ,
\qquad   V^\mu_\perp =  n_i^\mu  v^i  \,   .
\ee

\begin{figure}
\begin{center}
\includegraphics[height=6cm]{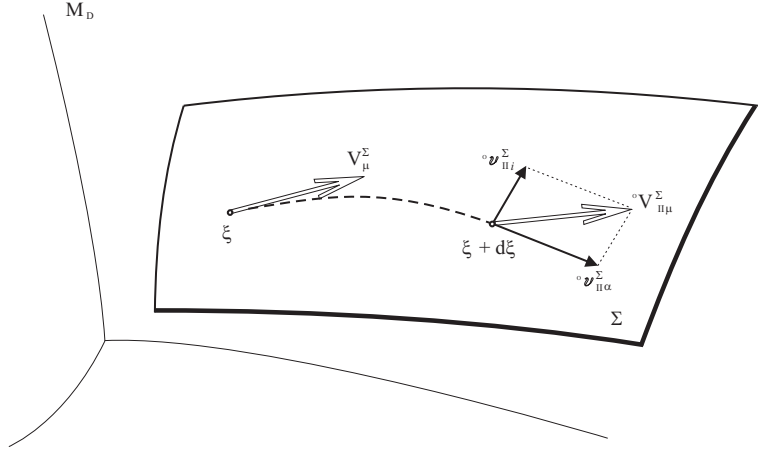}
\end{center}
\vspace*{-.5cm} \caption{{\it Definition of the induced connection}
        ${}^\circ \Gamma^\beta_{\alpha \gamma }$  {\it and SFF}
        ${}^\circ  b_{i \alpha \beta}$
        {\it from the parallel transport of the world-sheet tangent covector.
        Here} ${}^\circ  v_{\parallel \alpha}^\Sigma \equiv v_\alpha  +
        {}^\circ \Gamma^\beta_{\alpha \gamma }
        v_\beta \, d \xi^\gamma \,$ {\it and} ${}^\circ v^\Sigma_{\parallel i} \equiv -
{}^\circ b_{i \alpha \beta} \, v^\alpha d \xi^\beta $. }%
\end{figure}

Let us perform the parallel transport of the covector $V_\mu^\Sigma$ along
world-sheet line, from the point $\xi^\alpha$ to the point $\xi^\alpha + d
\xi^\alpha$, using the space-time connection ${}^\circ \Gamma^\nu_{\mu \rho}$
(Fig. 2). We obtain the covector
\be
{}^\circ V^\Sigma_{\parallel \mu} = V^\Sigma_\mu + {}^\circ \delta
V^\Sigma_\mu \, ,  \leq{ptv}
\ee
where
\be
{}^\circ \delta V_\mu^\Sigma ={}^\circ \Gamma^\nu_{\mu \rho} V_\nu^\Sigma d x^\rho ={}^\circ \Gamma^\nu_{\mu \rho}
V_\nu^\Sigma \partial_\gamma x^\rho d \xi^\gamma \,  .
\ee

In the local basis, at the point $\xi + d \xi$, its world-sheet projection
has the form
\be
{}^\circ v_{\parallel \alpha}^\Sigma  = \partial_\alpha x^\mu (\xi + d \xi) {}^\circ
V^\Sigma_{\parallel \mu}  \,  .
\ee

Let us introduce the world-sheet {\bf induced connection}, ${}^\circ
\Gamma^\beta_{\alpha \gamma }$, which defines the rule for the parallel
transport of the world-sheet covector $v_\alpha$, along the same world-sheet
line, to the projection ${}^\circ v_{\parallel \alpha}^\Sigma$. So, we have by
definition
\be
{}^\circ v_{\parallel \alpha}^\Sigma  = v_\alpha +{}^\circ \delta v_\alpha
\,  , \qquad  {}^\circ \delta v_\alpha ={}^\circ \Gamma^\beta_{\alpha \gamma }
v_\beta d \xi^\gamma \,  .  \leq{ptw}
\ee
It produce the expression for the induced connection
\be
{}^\circ \Gamma^\gamma_{\alpha \beta} = G^{\gamma \delta} \partial_\delta x^\mu G_{\mu \nu}
({}^\circ \Gamma_{\rho \sigma}^\nu   \partial_\alpha x^\rho \partial_\beta x^\sigma
+  \partial_\beta  \partial_\alpha x^\nu )=
G^{\gamma \delta} \partial_\delta x^\mu   G_{\mu \nu}
{}^\circ D_\beta \partial_\alpha x^\nu   \,  , \leq{icn}
\ee
where ${}^\circ D_\alpha V^\mu = \partial_\alpha x^\nu {}^\circ D_\nu V^\mu$
is space-time covariant derivative along world-sheet direction.

The orthogonal projection of the covector ${}^\circ V^\Sigma_{\parallel
\mu}\,$, defines the {\bf second fundamental form}, ${}^\circ  b_{i \alpha
\beta}\,$, trough the equation
\be
n_i^\mu (\xi+ d \xi) {}^\circ V^\Sigma_{\parallel \mu}
\equiv {}^\circ v^\Sigma_{\parallel i} = - {}^\circ b_{i \alpha
\beta}v^\alpha d \xi^\beta  \,   ,
\ee
or explicitly
\be
{}^\circ b_{i \alpha \beta} = n_i^\mu \,  G_{\mu \nu}\, {}^\circ \! D_\beta  \partial_\alpha x^\nu
 = -  \partial_\alpha x^\nu G_{\mu \nu} {}^\circ \! D_\beta  n_i^\mu         \,  .    \leq{sff2}
\ee
The SFF define the extrinsic geometry.

So, the induced connection and the SFF are the world-sheet and normal
projections respectively, of the space-time covariant derivative of
$\partial_\alpha x^\mu $. Consequently, we have the generalization of the
Gauss-Weingarten equation
\be
{}^\circ \! D_\beta  \partial_\alpha x^\mu = {}^\circ \Gamma^\gamma_{\alpha \beta}
\partial_\gamma x^\mu  + {}^\circ
b^i_{\alpha \beta} \, n_i^\mu      \,  .   \leq{scdd}
\ee

\subsection{Relations between space-time and world-sheet features}

In analogy with the general rule for connection decomposition, we
can decompose the induced connection and SFF. With the help of
\eq{cde} we have
\be
{}^\circ D_\beta \partial_\alpha x^\mu = D_\beta \partial_\alpha x^\mu
+  {}^\circ K^\mu{}_{\nu \rho}  \partial_\alpha x^\nu  \partial_\beta x^\rho  \,   . \leq{dscd}
\ee
The world-sheet projection of the last equation, produces the decomposition of
the induced connection
\be
{}^\circ \Gamma_{\gamma ,\alpha \beta} = \Gamma_{\gamma ,\alpha \beta} +
{}^\circ K_{\gamma \alpha \beta}   \,   ,     \leq{cdews}
\ee
in terms of induced Christoffel connection and {\bf induced torsion}
\be
\Gamma^\gamma_{\alpha \beta} = G^{\gamma \delta} \partial_\delta x^\mu \, G_{\mu \nu}
\, D_\beta \partial_\alpha x^\nu    \,  , \leq{ich}
\ee
\be
{}^\circ T_{\alpha \beta \gamma} ={}^\circ T_{\mu \rho \sigma} \partial_\alpha x^\mu
\partial_\beta x^\rho \partial_\gamma x^\sigma = {}^\circ \Gamma_{\alpha, \beta \gamma} -
{}^\circ \Gamma_{\alpha, \gamma \beta}   \,  ,   \leq{it}
\ee
where ${}^\circ K_{\gamma \alpha \beta} = \fr{1}{2} {}^\circ T_{\{\beta \gamma
\alpha \}}$ is the induced contortion. Note that the torsion is a
tensors, so that the corresponding relation does not have non-homogeneous
term.

The orthogonal projection of \eq{dscd} produces the decomposition of the SFF
\be
{}^\circ b_{i \alpha \beta} = b_{i \alpha \beta} + {}^\circ \! K_{i \alpha
\beta}  \,  ,
\ee
where we used the notations
\be
b_{i \alpha \beta} = n_i^\mu  G_{\mu \nu}  D_\beta  \partial_\alpha x^\nu  \,   ,  \qquad
{}^\circ T_{i \alpha \beta} =
{}^\circ T_{\mu \rho \sigma} n_i^\mu \partial_\alpha x^\rho  \partial_\beta x^\sigma
      \,   ,  \leq{sffCh}
\ee
and similarly as before ${}^\circ K_{i \alpha \beta} = \fr{1}{2} {}^\circ
T_{\{\beta i \alpha \}}$.

When the torsion is present, the SFF is not symmetric in $\alpha,\beta$
indices. Similarly as in the case of the connection, we can write
\be
{}^\circ b_{i \alpha \beta} = {}^\circ b_{i ( \alpha \beta)} +
\fr{1}{2} {}^\circ T_{i \alpha \beta} \,   .
\ee

Starting with the definition of the space-time covariant derivatives along
world-sheet direction, ${}^\circ D V^\mu = V^\mu ( \xi + d \xi) -
V^\mu_{\parallel}$, we can obtain the relations
\be
{}^\circ D_\alpha V^\mu_\Sigma = {}^\circ \nabla_\alpha v^\beta \, \partial_\beta x^\mu +
n_i^\mu \, {}^\circ b^i_{\beta \alpha} v^\beta  \,  ,   \leq{Ge}
\ee
\be
({}^\circ D_\alpha V_\mu) \partial_\beta x^\mu = {}^\circ \nabla_\alpha
v_\beta - v^i \, {}^\circ b_{i \beta \alpha} \,  ,  \leq{rcd}
\ee
which we will  need later.

\section{World-sheet as a minimal surface}
\setcounter{equation}{0}

In this section we will present geometrical interpretation of the field
equations in the absence of antisymmetric tensor field $B_{\mu \nu}$.

\subsection{Mean extrinsic curvature}

In the torsion free case, the SFF is symmetric in the world-sheet indices and
its properties are well known in the literature. Here, we will generalize it
for the spaces with Minkowski signature.

As usual, a curve is parametrized by its length parameter $s$, so that the
unit tangent vector is $t^\mu = {d x^\mu \over ds}$. If the curve lies in the
world-sheet, we have $t^\mu =  t^\alpha \partial_\alpha x^\mu$, with $t^\alpha
= {d \xi^\alpha \over ds}$. Let us denote by $P_i$, the 2-dimensional plane
spanned by the tangent vector $t^\mu$ and the unit world-sheet normal
$n_i^\mu$. Then, $\ell_i = \Sigma \bigcap P_i$ is the $i$-th normal section of
the world-sheet $\Sigma$. The curvature of the normal section $\ell_i$, as a
space-time curve, is
\be
{}^\circ k_i = {}^\circ D_s t^\mu G_{\mu \nu} n^\nu_i  \,  ,
\ee
where ${}^\circ D_s t^\mu \equiv {\dot x}^\nu {}^\circ D_\nu t^\mu$, is  the
covariant derivative along the curve. It produces the following expression
\be
{}^\circ k_i = {}^\circ  b_{i \beta \alpha}\, t^\alpha t^\beta =
{}^\circ  b_{i (\beta \alpha)}\, t^\alpha t^\beta = { {}^\circ
b_{i (\beta \alpha)} d \xi^\alpha d \xi^\beta \over G_{\alpha
\beta} d \xi^\alpha d \xi^\beta} \,  .       \leq{cns}
\ee
The curvature ${}^\circ k_i$ depends on the direction of the tangent vector $
t^\alpha$ and only on the symmetric part of the SFF.

In Euclidean spaces, the maximal and minimal values of ${}^\circ \! k_i$ are
the principal curvatures. The corresponding directions, defined by $t^\alpha$,
are the principal directions. The principal curvatures are eigenvalues of the
SFF and corresponding eigenvectors define the principal directions.

In the present case with Minkowski space-time, the curvature ${}^\circ \!
k_i$,  \eq{cns}, is divergent in the light-cone directions. Consequently, the
extremely values do not exist. Still, we can obtain the necessary information
from the eigenvalue problem
\be
({}^\circ \! b^i_{\alpha \beta} -  {}^\circ \! \kappa^i G_{\alpha
\beta}) v^\beta_i = 0  \,   .           \leq{ssp}
\ee

The eigenvalues of the quadratic forms ${}^\circ  b^i_{\alpha \beta}$ with
respect to the metric $G_{\alpha \beta}$, we define as a principal curvatures
in Minkowski space-time. They are the solutions of the condition
$\det({}^\circ \! b^i - {}^\circ \!\kappa^i G)_{\alpha \beta} = 0$, or
explicitly
\be
{}^\circ \! \kappa^i_{0,1} =  {}^\circ \! H^i \pm \sqrt{ ({}^\circ \! H^i)^2 -  {}^\circ \!
K^i} \,    .
\ee
Here
\be
{}^\circ \! H^i = \fr{1}{2} G^{\alpha \beta} {}^\circ \! b^i_{\alpha \beta} \,
 =  \fr{1}{2} ({}^\circ \! \kappa^i_0 + {}^\circ
\! \kappa^i_1)  \,   ,  \leq{aec}
\ee
is the trace of the SFF, known in the literature as {\bf mean extrinsic
curvature} and ${}^\circ \! K^i = { \det {}^\circ \! b^i_{\alpha
\beta} \over \det G_{\alpha \beta}}= {}^\circ \! \kappa^i_0 {}^\circ \!
\kappa^i_1 $ (no summation over $i$) is Gauss curvature.

The surface defined by the equation ${}^\circ \! H^i = 0$ is {\bf minimal
surface}. The name stems from the fact that in the Riemann space-time the
equation $H_i = 0$ define the surface of minimal area $P_2 = \int d^2 \xi
\sqrt{-\det G_{\alpha \beta}}\,$, for the fixed boundary. In fact, for this
surface the first variation of the area vanishes
\be
\delta P_2 = -2  \int d^2 \xi \sqrt{-\det G_{\alpha \beta}} \, H^i n^\mu_i G_{\mu \nu} \, \delta x^\nu
= 0  \,  .
\ee

\subsection{Local properties of the world-sheet embedded in Riemann space-time}

In the absence of antisymmetric field, $B_{\mu \nu}=0$, the equations of
motion for the action \eq{ac},  survive in the simpler form
\be
[x^\mu]  \equiv \, \nabla_- \partial_+ x^\mu + \Gamma_{\rho \sigma}^\mu
\partial_+ x^\rho  \partial_- x^\sigma =0    \, ,  \qquad
[h^\pm]  \equiv G_{\mu \nu}  \partial_\pm  x^\mu \partial_\pm x^\nu =0 \,   ,
\ee
where $\Gamma^\mu_{\rho \sigma}$ is the Christoffel connection. In this case
the string does not see the torsion and it feels the target space as {\it
Riemann space-time} of general relativity \cite{MB,HMMN}.

The world-sheet and orthogonal projections of the equation $[x^\mu]$,  obtain
the forms
\be
g^{\alpha \beta} (\con^\gamma_{ \alpha \beta } - \Gamma^\gamma_{ \alpha \beta })=0  \,   , \qquad
g^{\alpha \beta} b^i_{\beta \alpha} = 0  \,  .   \leq{ic}
\ee
Here, $\Gamma_{\alpha \beta}^\gamma$ is induced world-sheet Christoffel
connection \eq{ich} and $b^i_{\alpha \beta}$ is corresponding SFF \eq{sffCh}.
By $\con^\gamma_{ \alpha \beta }$ we denote intrinsic world-sheet connection.

The $[h^\pm]$ equation can be written in the form
\be
G_{\pm \pm} \equiv e_\pm^\alpha e_\pm^\beta  G_{\alpha \beta} = 0     \,  , \leq{imtem}
\ee
where $G_{\alpha \beta} = G_{\mu \nu} \partial_\alpha x^\mu \partial_\beta
x^\nu $ is world-sheet induced metric. Because in the light-cone frame, the
intrinsic metric is off-diagonal, $g_{\pm \pm} =0$, we have $G_{a b} = \lambda
g_{a b}, \,\, (a,b \in \{+, -\})$, or equivalently
\be
G_{\alpha \beta } = \lambda g_{\alpha \beta}     \,      .  \leq{icc}
\ee
Multiplying the last equation with $g^{\alpha \beta}$, we obtain the
expression $ \lambda = \fr{1}{2}g^{\alpha \beta}G_{\alpha \beta }$, so that
$\sqrt{-G} = \sqrt{-g} \fr{1}{2}g^{\alpha \beta}G_{\alpha \beta }$. The last
relation connects the Polyakov and Nambu-Goto expressions for the string
action. We can rewrite \eq{icc} in the form
\be
{G_{\alpha \beta } \over \sqrt{-G}} =  {g_{\alpha \beta } \over \sqrt{-g}} \,  ,
\ee
so that, because of the conformal invariance, only the metric densities are
related.

From \eq{icc} follows a relation between intrinsic connection $\con^\gamma_{
\alpha \beta }$ and induced one $\Gamma^\gamma_{ \alpha \beta }$ which is also
a solution of the first equation \eq{ic}. Therefore, both intrinsic metric
tensor and connection, are equal to the induced ones from the space-time, up
to the conformal factor $\lambda$. This is expected result, because of the
conformal invariance of the action.

With the help of  \eq{icc} the second equation \eq{ic} becomes
\be
H^i \equiv \fr{1}{2}G^{\alpha \beta} b^i_{\beta \alpha} = 0 \,  .
\ee
Therefore, the vanishing of all MECs, are local properties of minimal
world-sheet in Riemann space-time.

Out of the initial, $D$ components of $[x^\mu]$ equations, two define intrinsic
connection in terms of the induced one and $D-2$ turn all MECs to zero.

\section{ World-sheet as a CT-dual surface}

Let us stress that above considerations are torsion independent, because the
antisymmetric part of the SFF disappears from  \eq{cns} and  \eq{aec}. In this
section we are going to include the torsion contribution and generalize above
results.

\subsection{CT-duality between mean extrinsic curvature and extrinsic mean torsion}
\setcounter{equation}{0}

Let us first generalize the eigenvalue problem, and then offer its geometrical
interpretation. We introduce the {\bf dual eigenvalue } problem, such that
linear transformation of the vector $v^\alpha$ with matrix  ${}^\circ \!
b^i_{\alpha \beta}$ is proportional to the two dimensional dual vector
${}^\ast v_\alpha = \sqrt{-G_2} \, \varepsilon_{\alpha \beta} \, v^\beta \,\,
(G_2 = \det G_{\alpha \beta})$
\be
{}^\circ \! b^i_{\alpha \beta} \, v^\beta = {}^\ast \kappa^i \, {}^\ast v_\alpha
\,    ,
\ee
or equivalently
\be
\left( {}^\circ \! b^i_{\alpha \beta} -  {}^\ast \kappa^i \, \varepsilon_{\alpha
\beta} \sqrt{-G_2} \right) v^\beta_i = 0  \,   .           \leq{asp}
\ee
It is similar to \eq{ssp}, but for the completeness we also need the
eigenvalues of the quadratic forms ${}^\circ  b^i_{\alpha
\beta}$ with respect to the antisymmetric tensor $\varepsilon_{\alpha \beta}
\sqrt{-G_2}$.

We can formulate the dual eigenvalue problem  \eq{asp}, as an ordinary
eigenvalue problem $({}^\ast b^i_{\alpha \beta}  - {}^\ast \kappa^i G_{\alpha
\beta} ) v^\alpha_i = 0$, if we introduce the dual SFF
\be
{}^\ast b^i_{\alpha \beta} = {G_{\alpha \gamma}
\varepsilon^{\gamma\delta} \over  \sqrt{-G_2} } \, {}^\circ \! b^i_{\delta
\beta} \,  .
\ee

The solutions of the condition $\det( {}^\ast \! b^i -  {}^\ast \kappa^i \,
G)_{\alpha \beta} = 0$, have the form
\be
{}^\ast \kappa^i_{0,1} =  {}^\ast H^i \pm \sqrt{({}^\ast H^i)^2 + {}^\circ \!
K^i} \,    .
\ee
In analogy with the previous case, we will call them dual principal curvatures,
and the variable
\be
{}^\ast H^i = \fr{1}{2} ({}^\ast \kappa^i_0 + {}^\ast \kappa^i_1) =
\fr{1}{2} G^{\alpha \beta} {}^\ast  b^i_{\alpha \beta} =
\fr{1}{2} {\varepsilon^{\alpha \beta} \over \sqrt{-G_2}} {}^\circ \! b^i_{\beta \alpha} =
\fr{1}{4} {\varepsilon^{\alpha \beta} \over \sqrt{-G_2}}
{}^\circ \! T^i_{\beta \alpha}     \,  ,          \leq{dmc}
\ee
the {\bf dual mean extrinsic curvature}. Here ${}^\circ \! K^i =
{ \det {}^\circ \! b^i_{\alpha \beta} \over \det G_{\alpha
\beta}}= {}^\ast \kappa^i_0 {}^\ast \kappa^i_1 $ (no summation over $i$) is the
same Gauss curvature as before.

Let us turn to the geometrical meaning of the DMEC. In the case when $t^\mu_1$
and $t^\mu_2$ are world-sheet tangent vectors (Fif. 1) (note that all
geodesics $\ell_1$, $\ell_2$, $\tilde{\ell}_1$ and $\tilde{\ell}_2$ still
could be space-time curves) we can rewrite  \eq{cod} in the form
\be
{}^\circ T^\mu  \equiv  {x^\mu(D_2) - x^\mu(D_1) \over 2 d P_{12}} =
{\varepsilon^{\beta \alpha} \over 4 \sqrt{-G_2}}\,
{}^\circ T^\mu_{\rho \sigma} \partial_\alpha x^\rho  \partial_\beta x^\sigma  \,    .
\ee
Here, $ d P_{12} = \sqrt{-G_2} \det( {\partial \xi^\alpha \over \partial s_r})
d \ell_1 d \ell_2$ is area of the parallelogram, spanned by the vectors
$\ell_1^\mu = t_1^\mu d \ell_1$ and $\ell_2^\mu = t_2^\mu d \ell_2\,$. We can
conclude that ${}^\circ T^\mu$ does not depend on the directions $t^\mu_1$ and
$t^\mu_2$, and on the lengths $d \ell_1$ and $d \ell_2$. So, we will refer to
this variable as the {\bf mean torsion}. Its world-sheet projection is induced
mean torsion
\be
{}^\circ T_\gamma  = {}^\circ T^\mu G_{\mu \nu} \partial_\gamma x^\nu =
 {\varepsilon^{\beta \alpha} \over 4 \sqrt{-G_2}}\, {}^\circ T_{\gamma \alpha \beta}   \,
 . \leq{mt}
\ee
Its normal projection is the {\bf extrinsic mean torsion} (EMT)
\be
{}^\circ T_i  = {}^\circ T^\mu G_{\mu \nu} n_i^\mu =
{\varepsilon^{\beta \alpha} \over 4 \sqrt{-G_2}} \, {}^\circ T_{i \alpha \beta}
=  {}^\ast H_i   \,  ,  \leq{emt}
\ee
which is exactly the same variable as DMEC, defined above in \eq{dmc}.

We define the {\bf CT-duality} ({\bf C}urvature-{\bf T}orsion duality), which
maps SFF to dual SFF, ${}^\circ b^i_{\alpha \beta} \to {}^\ast b^i_{\alpha
\beta}$, and interchanges the role played by the symmetric and the
antisymmetric parts of the SFF. Consequently, under CT-duality MEC maps to
DMEC, allowing the exchange of the mean extrinsic curvature and extrinsic mean
torsion.

The self-dual and self-antidual configurations
\be
{}^\circ H^i = \pm  {}^\ast H^i  \,   , \qquad \Leftrightarrow  \qquad {}^\circ H^i = \pm  {}^\circ T^i  \,   ,   \leq{sdr}
\ee
correspond to the following conditions on the SFF
\be
(G^{\alpha \beta} \mp {\varepsilon^{\alpha \beta} \over
\sqrt{-G_2}}) {}^\circ \! b^i_{\beta \alpha} = 0  \,  .   \leq{sasd}
\ee
The equations  \eq{sdr} and \eq{sasd} define {\bf CT-dual (antidual)
surfaces}. In the torsion free case, they turn to the standard minimal surface
condition, ${}^\circ H^i= 0$.

\subsection{Local properties of the world-sheet embedded in Riemann-Cartan space-time}

Let us applay the results of the previous subsection to the action \eq{ac}
when both metric tensor $G_{\mu \nu}$ and the antisymmetric field $B_{\mu
\nu}$ are present, so that we have complete equations of motion \eq{lJ} and
\eq{lh}. In this case, the string feels the target space as {\it
Riemann-Cartan space-time}.

As well as in the case of Riemann space-time, the same relation between metric
tensors, \eq{icc}, follows from the $[h^\pm]$ equation. We can rewrite the
$[x^\mu]$ equation in the form
\be
g^{\alpha \beta} ( D_\beta \partial_\alpha x^\mu - \con^\gamma_{\alpha
\beta} \partial_\gamma x^\mu) = { \varepsilon^{\alpha \beta} \over
\sqrt{-g}}B^\mu_{\alpha \beta}  \,  .
\ee
Its world-sheet projection produces
\be
g^{\alpha \beta} (\con^\gamma_{\alpha \beta} - \Gamma^\gamma_{ \alpha \beta})
= - { \varepsilon^{\alpha \beta} \over \sqrt{-g}}B^\gamma_{\alpha
\beta}    \,  ,   \leq{cei}
\ee
where $B^\gamma_{\alpha \beta}$ is the world-sheet torsion, induced from the
target space. Because it is totally antisymmetric, in two dimensions it
vanishes, $B^\gamma_{\alpha \beta} =0$. So, we obtain the same first equation
\eq{ic}, as in the case of Riemann space-time. Again, both two dimensional
intrinsic metric tensor and two dimensional intrinsic connection, up to the
conformal factor, are induced from the target space. Note that the world-sheet
is torsion free while the space-time is not.

The orthogonal projection of the $[x^\mu]$ equation takes the form
\be
g^{\alpha \beta} b_{i \alpha \beta} =  { \varepsilon^{\alpha \beta} \over
\sqrt{-g}} B_{i \alpha \beta}  \, ,   \leq{dem}
\ee
or equivalently
\be
(g^{\alpha \beta} - { \varepsilon^{\alpha \beta} \over \sqrt{-g}})
b_{- i \beta \alpha} = 0   \, .   \leq{dem1}
\ee

With the help of \eq{icc}, we can rewrite it in terms of induced metric
\be
( G^{\alpha \beta} - { \varepsilon^{\alpha \beta} \over
\sqrt{-G_2}}) b_{- i \beta \alpha} = 0 \,  ,  \qquad
\Leftrightarrow    \qquad   H_i =  {}^\ast H_{- i}     \,      . \leq{sdrc}
\ee
Therefore, according to the local property, the world-sheet embedded in
Riemann-Cartan space-time is {\it CT-dual surface} and the field equations
have a form of CT-duality.

As a consequence of the relation $\Gamma^\mu_{+ \rho \sigma} = \Gamma^\mu_{-
\sigma \rho}$, we have $b_{+  i \alpha \beta} = b_{- i \beta \alpha}$, so
that with respect to $\Gamma^\mu_{+ \rho \sigma}$ the world-sheet is CT-antidual
surface.

\section{Conclusions}

In this paper, we investigated the bosonic string propagating in the
nontrivial background. We described the classical field equations of  in term
of world-sheet local properties valid in the neighborhood of every given
point.

We started with geometry of the surface embedded into
Riemann-Cartan space-time. We clarified the meaning of MEC in
Minkowski space-time ${}^\circ H^i$ (see \eq{aec}), and
introduced the concept of DMEC ${}^\ast H^i$ \eq{dmc} as
orthogonal projection of the mean torsion (see \eq{mt} and
\eq{emt}). We defined CT-duality which maps MEC to DMEC. The
presence of torsion generalize the equation of minimal surfaces.
Instead of the standard equation ${}^\circ H_i = 0$, we introduced
CT-dual (antidual) surface defined by the self-duality
(self-antiduality) conditions, ${}^\circ H_i = \pm {}^\ast H_i$.

Then we considered the equations of motion \eq{lJ}-\eq{lh}. As a consequence
of the second equation, the intrinsic metric tensor is equal to the induced
one up to the conformal factor $\lambda$, because the theory is conformally
invariant.

The first equation, which has been obtained by variation with respect to
$x^\mu$, have $D$ components. Two of them determine the contracted intrinsic
connection in terms of the corresponding induced one. They are not
independent, because they follow from the $[h^\pm]$ equation.

The other $D-2$ are of the form $ H_i = {}^\ast H_{- i}$, where $ H_i$ is  MEC
and ${}^\ast H_{- i}$ is DMEC. They define world-sheet as CT-dual surface of
the Riemann-Cartan space-time. In the particular case,
---the vanishing torsion --- the field equations turn to the equations of
minimal world-sheet, $H_i = 0$, of the Riemann space-time.

The dilaton field is origin of nonmetricity \cite{BS3}. It broke the conformal
invariance and produce additional field equation with respect to conformal
part of world-sheet metric. The analysis of the dilaton contribution to the
local properties of the world-sheet, which is technically more complicated,
has been investigated in \cite{BS3}.

\appendix 

\section{World-sheet geometry}
\setcounter{equation}{0}

It is useful to parameterize the intrinsic world-sheet metric tensor
$g_{\alpha \beta }$, with the light-cone variables  $(h^+,h^-, F)$ (see the
papers \cite{BS1,BPS,SM})
\be
g_{\alpha \beta} =e^{2F} {\hat g}_{\alpha \beta}=
\fr{1}{2}e^{2F}\pmatrix{ -2h^-h^+    &  h^-+h^+ \cr
           h^-+h^+    &  -2      \cr }\,  .               \leq{g}
\ee
The world-sheet interval
\be
ds^2 = g_{\alpha \beta} d \xi^\alpha d \xi^\beta = 2 d \xi^+ d
\xi^-  \,   ,
\ee
can be expressed in terms of the variables
\be
d \xi^\pm = { \pm 1 \over \sqrt{2}} e^F ( d \xi^1 - h^\pm d \xi^0)
= e^F d {\hat \xi}^\pm  = e^\pm{}_\alpha  d \xi^\alpha \,    .
\ee
The quantities $ e^\pm{}_\alpha$ define the light-cone one form basis,
$\theta^\pm = e^\pm{}_\alpha d \xi^\alpha$, and its inverse define the tangent
vector basis, $e_\pm = e_\pm{}^\alpha \partial_\alpha = \partial_\pm$. We will
use the relations
\be
\eta^{ab} e_a{}^\alpha e_b{}^\beta = e_+{}^\alpha e_-{}^\beta +
e_-{}^\alpha e_+{}^\beta = g^{\alpha \beta} \,  , \qquad
\varepsilon^{ab} e_a{}^\alpha e_b{}^\beta = e_+{}^\alpha
e_-{}^\beta - e_-{}^\alpha e_+{}^\beta = {\varepsilon^{\alpha
\beta} \over \sqrt{-g}} \,  , \leq{mtast}
\ee
where $a,b \in \{+,-\} $.

In the tangent basis notation, the components of the arbitrary vector
$V_\alpha$ have the form
\be
V_{\pm}=e^{-F} {\hat V}_\pm =e_{\pm}{}^\alpha V_\alpha =
{\sqrt{2} e^{-F} \over h^- -h^+} (V_0+h^{\mp}V_1)\,  .     \leq{vec}
\ee
The world-sheet covariant derivatives on tensor $X_n$ are
\be
\nabla_\pm X_n = (\pd_{\pm} +n \con_{\pm}) X_n   \,  ,        \leq{wscd}
\ee
where the number $n$ is sum of the indices, counting index $+$ with $1$ and
index $-$ with $-1$, and
\be
\con_{\pm} =e^{-F}({\hat \con}_\pm \mp {\hat \partial}_\pm F) \,  , \qquad
{\hat \con}_\pm =\mp {\sqrt{2}\over h^- -h^+} h^{\mp \prime} \,  ,
\ee
are two dimensional Riemannian connections.

\end{document}